\documentclass[aps,pre,preprint,superscriptaddress,showkeys,showpacs,amssymb]{revtex4}
\usepackage{graphicx}
\usepackage{graphics}
\usepackage{amsmath}
\usepackage{tabularx}
\usepackage{color}
\usepackage{doi}
\usepackage{textcomp}
\usepackage{dcolumn}
\usepackage{bm}
\newcolumntype{Y}{>{\centering\arraybackslash}X}

\begin{document}

\title{Modified Epidemic Diffusive Process on the Apollonian Network}

\author{D. S. M. Alencar}
\affiliation{Departamento de F\'{\i}sica, Universidade Federal do Piau\'{i}, 57072-970, Teresina - PI, Brazil}
\author{A. Macedo-Filho}
\affiliation{Departamento de F\'{i}sica, Universidade Estadual do Piau\'{i}, 64002-150, Teresina - PI, Brazil}
\author{T. F. A. Alves}
\affiliation{Departamento de F\'{\i}sica, Universidade Federal do Piau\'{i}, 57072-970, Teresina - PI, Brazil}
\author{G. A. Alves}
\affiliation{Departamento de F\'{i}sica, Universidade Estadual do Piau\'{i}, 64002-150, Teresina - PI, Brazil}
\author{R. S. Ferreira}
\affiliation{Departamento de Ci\^{e}ncias Exatas e Aplicadas, Universidade Federal de Ouro Preto, 35931-008, Jo\~{a}o Monlevade - MG, Brazil}
\author{F. W. S. Lima}
\affiliation{Departamento de F\'{\i}sica, Universidade Federal do Piau\'{i}, 57072-970, Teresina - PI, Brazil}


\date{Received: date / Revised version: date}

\begin{abstract}

We present an analysis of an epidemic spreading process on the Apollonian network that can describe an epidemic spreading in a non-sedentary population. The modified diffusive epidemic process was employed in this analysis in a computational context by means of the Monte Carlo method. Our model has been useful for modeling systems closer to reality consisting of two classes of individuals: susceptible (A) and infected (B). The individuals can diffuse in a network according to constant diffusion rates $D_{A}$ and $D_{B}$, for the classes A and B, respectively, and obeying three diffusive regimes, i.e., $D_{A}<D_{B}$, $D_{A}=D_{B}$ and $D_{A}>D_{B}$. Into the same site $i$, the reaction occurs according to the dynamical rule based on Gillespie's algorithm. Finite-size scaling analysis has shown that our model exhibit continuous phase transition to an absorbing state with a set of critical exponents given by $\beta/\nu=0.66(1)$, $1/\nu=0.46(2)$, and $\gamma/\nu=-0.24(2)$ common to every investigated regime. In summary, the continuous phase transition, characterized by this set of critical exponents, does not have the same exponents of the Mean-Field universality class in both regular lattices and complex networks.

\end{abstract}


\maketitle

\section{\label{sec:level1}Introduction}

Nowadays, we have passed by an infectious disease named COVID-19, which was considered a threat to global public health. This epidemic process caused acute pneumonia in patients around the world, being widely investigated in many areas of the science\cite{zaher2020,sun2020,lu2020,zhang2020}. Based on this, we report that epidemic processes have been widely studied over the last years, for instance, by the physicists' community. Thus, many models were created and applied to mimic and to understand such epidemic processes, including the Susceptible-Infected-Susceptible (SIS) model \cite{bartlett1956,castellano2020,mota2018,macedo2018}, Susceptible-Infected-Recovered (SIR) model \cite{bailey1957,santos2020,alencar2020}, The Contact Process (CP) model \cite{harris1974,lfsilva2013,marro1999,almeida2016}, Diffusive Epidemic Process (DEP) model \cite{kree1989,wijland1998,fulco2001,maia2007,amfilho2010}, among others.

These processes belong to the class of non-equilibrium systems and usually display a continuous phase transition to an absorbing state. To investigate these processes, we have a wide theoretical field based on analytical techniques such as Mean Field (MF) theory and Renormalization Group (RG) theory, besides computational methods that rely on Monte Carlo (MC) method simulations that were developed in the last years \cite{hinrichsen2000}. In this last case, major support has been coming from technological advances. Thereby, through the use of computational tools, we can model real situations and make predictions, in a short time, that are useful to human society.

Over the last years, the DEP model has been useful for modeling systems closer to reality by inclusion of a non-sedentary population in order to describe a reaction-diffusion process. This model consists of two classes of individuals, susceptible (A) and infected (B), diffusing on a lattice obeying three diffusive regimes, i.e., $D_{A}<D_{B}$, $D_{A}=D_{B}$, and $D_{A}>D_{B}$, where $D_{A}$ and $D_{B}$ are constant probabilities in the [$0,1$] interval. It is worth mentioning that Kree et al. analyzed the DEP model for the first time investigating an ecological system, describing the effect of pollution on a population that is on the brink of extinction, aiming to understand ecological catastrophes, by using RG theory.

Thus, according to the RG analysis, the system presented a continuous phase transition to an absorbing state with critical exponents $\eta=-\epsilon/8$ to first order of $\epsilon=4-d$, $\nu_{\perp}=2/d$ and $z=2$ in all orders in $\epsilon$. Next, the Kree et al. model\cite{kree1989} was extended by van Wijland et al.\cite{wijland1998}, considering the $D_{A}>D_{B}$, $D_{A}=D_{B}$, and $D_{A}<D_{B}$ regimes. In this context, the RG theory predicted a continuous phase transition for the $D_{A}<D_{B}$ regime, defining a new universality class (WOH) with critical exponents given by $\nu_{\perp}=2/d$, $\eta=0$ and $z=2$ in all orders in $\epsilon$, and recovering the Kree et al. universality class (KSS) studied to $D_{A}=D_{B}$ regime. In addition, to the $D_{A}>D_{B}$ regime, they conjectured a discontinuous phase transition, however, MC simulations analysis with distinct algorithms are consistent with a continuous phase transition to all three regimes\cite{fulco2001,maia2007,amfilho2010}.

In this work, we performed a study about the DEP model coupled to Apollonian networks, a particular complex network model introduced by J. S. Andrade Jr. et al.\cite{jsandrade2005} in 2005, which is characterized by a degree distribution given by a power-law $P(k) \propto k^{-\gamma}$, where $k$ is the number of connected neighbors (degree), and $\gamma=1 + \ln{3}/\ln{2} \approx 2.585$ is the power-law exponent. In this context, according to the MF theory, the critical behavior of the system depends on the distribution degree $\gamma$. Thus, we can classify three $\gamma$ regions\cite{lfsilva2013,scferreira2011,jdnoh2009,rsferreira2011}, i.e.:
\begin{itemize}
\item $\gamma < 2$ there is no existing phase transition;
\item $2 < \gamma \leq 3$ with phase transition characterized by $\beta=1/(\gamma-2)$ and $\nu=(\gamma-1)/(\gamma-2)$ exponents, and 
\item $\gamma > 3$ with critical exponents $\beta=1$ and $\nu=2$.
\end{itemize}
Here, our main interest in networks with power-law distribution is the scale-free property, specifically the power-law exponent satisfying $2 < \gamma \leq 3$ interval, yielding a finite average degree scaling as $\langle k\rangle\approx k_c^{2-\gamma}$ (first moment) while the second (and higher) moment diverges as $k_c\approx k_c^{3-\gamma}$, where $k_c$ is the maximum degree present in the network. Besides that, scale-free networks are ubiquitous in nature and society\cite{cohen2010,barabasi2016,lijeros2001,mendes2009,almeida2013,price1965}.

In this context, our investigation consists in a modified DEP (MDEP) model study that aims to estimate the existence of phase transition in each $D_{A}>D_{B}$, $D_{A}=D_{B}$ and $D_{A}<D_{B}$ regimes, in addition to its critical exponents $\beta/\nu$, $1/\nu$, and $\gamma/\nu$. Thus, the main modification to the DEP usual definition was done on the reaction stage, by simulating the reaction process as a chemical reaction by using Gillespie algorithm\cite{gillespie2013,gillespie1976, gillespie1977}, in order to introduce a finite threshold in the scale-free networks\cite{tfaalves2020}. This algorithm allows to stochastically solve the differential equations like SIS model\cite{pastorsatorras2015} of a homogeneous population, not coupled to a network, and find the time evolution of the infected and susceptible compartments $A$ and $B$, respectively
\begin{eqnarray}
\frac{\mathrm{d}}{\mathrm{d}t} \rho_{A} &=& - \left(\mu_c\frac{\rho_{B}}{\rho}\right) \rho_{A} + \mu_r \rho_{B}, \nonumber \\
\frac{\mathrm{d}}{\mathrm{d}t} \rho_{B} &=& + \left(\mu_c\frac{\rho_{B}}{\rho}\right) \rho_{A} - \mu_r \rho_{B},
\label{sis-diffeq}
\end{eqnarray}
where $\mu_c$ is the infection rate, $\mu_r$ is the recovery rate and $\rho=\rho_{A}+\rho_{B}$ is the total density population. When coupling a population of random walkers to a network, one can make the density population $\rho_{A}(i)$ and $\rho_{B}(i)$ of node $i$ obeying the Eq.(\ref{sis-diffeq}), where the reactions take place in a time $t_\mathrm{max}$ while maintaining the discrete time diffusion. More details are shown in section \ref{sec:level2}.

This paper is organized as follows: in section \ref{sec:level2} we describe the Apollonian network and MDEP model and its dynamics. In section \ref{sec:level3} we present our simulation results and discussions about our proposed model. And finally,  in section \ref{sec:level4} we present our conclusions.

\section{\label{sec:level2}Apollonian network and Modified diffusive epidemic process}

\subsection{Apollonian networks}

\begin{figure}
\includegraphics[scale=0.4]{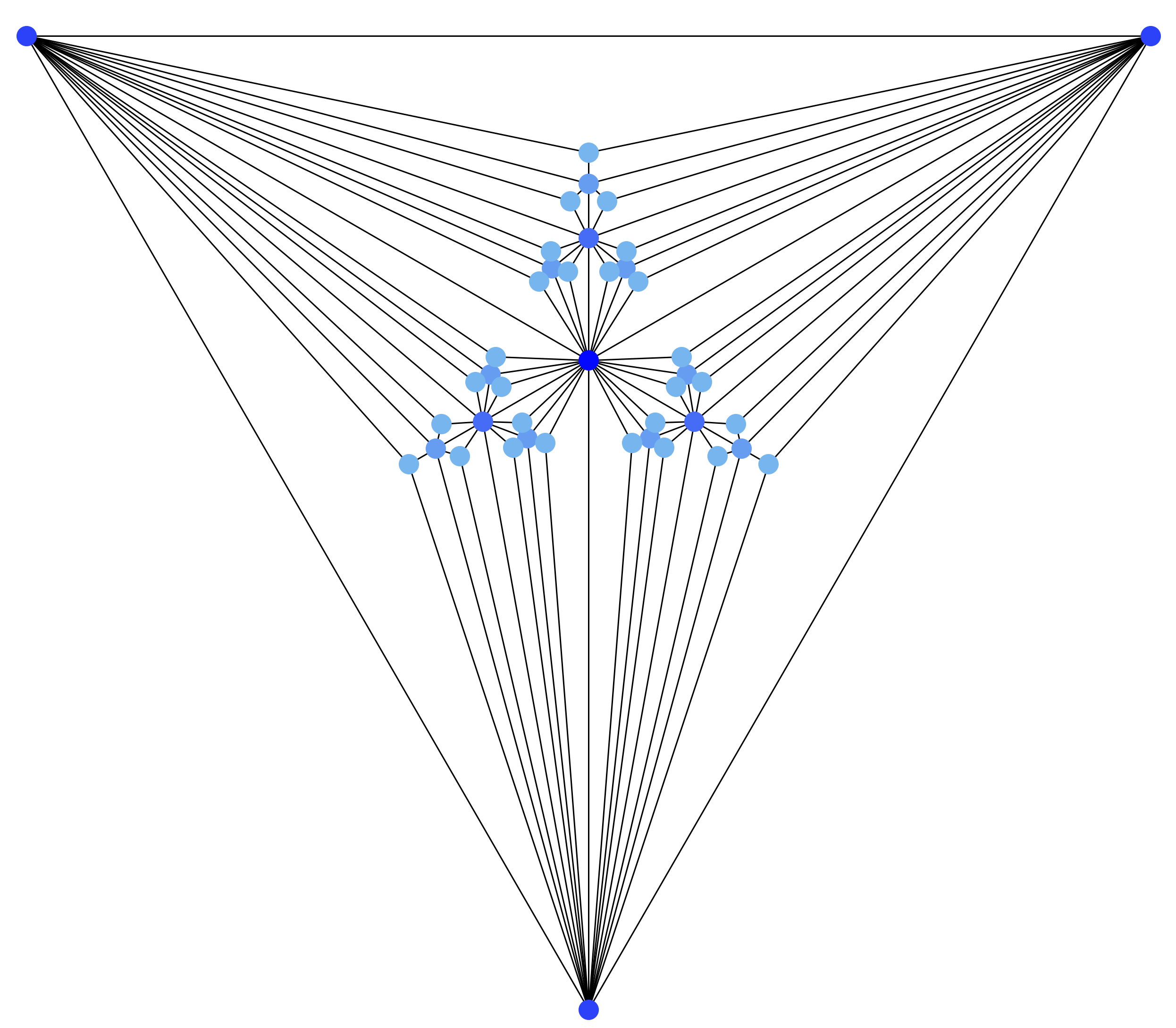}
\caption{(Color online). Apollonian network of fourth generation. This generation contains $N=43$ nodes.}
\label{fig:apn}
\end{figure}

The Apollonian network belongs to the complex network family, having its origin based on the problem of a space-filling packing of disks first proposed by the Greek mathematician Apollonius of Perga \cite{boyd1973}. To build it, we consider three equal radius disks touching each other and the space among them is filled by another disk that touches all the previous three disks. This procedure can be iterated by inserting smaller disks inside the space among any three touching disks. The network is formed by connecting the centers of the touching disks, obtaining a network that gives a triangulation that physically corresponds to the force network of a dense granular packing. The number of nodes $N$ at each generation $n=0,1,2,3,\dots$ can be found according to the relation 
\begin{equation}
N=3+(3^{n+1}-1)/2.
\end{equation}
As an example, Fig.(\ref{fig:apn}) displays an Apollonian network of fourth-generation with $N=43$ nodes. Next, it is worth mentioning that a detailed description of the Apollonian network building can be found in Ref.\cite{gaalves2020}.

In comparison to other complex network models, the Apollonian network is a particular case of a deterministic complex scale-free network embedded in a Euclidean lattice with space-filling and matching graph properties. Also, it displays small-world effect, which means that the average length of the shortest paths \textit{l} between two any nodes grows up slower than any positive power of the system size $N$. In addition, it presents a large clustering coefficient \textit{C}. Thus, since \textit{l} grows up logarithmically and \textit{C} tends to unity, the Apollonian network indeed exhibits a small-world effect. In this sense, the literature shows that the average length of the shortest paths is $l\propto [\textrm{ln}(N)]^{3/4}$ and clustering coefficient $C = 0.828$ in the limit of large $N$\cite{wattsstrogatz1998,jsandrade2005,jspkdoyde2005,lfsilva2013}.

\subsection{Modified diffusive epidemic process and implementation}

Here, we introduce the DEP simulation based on kinetic Monte-Carlo dynamic rules. The main modification introduced is relative to the reaction process. According to the usual definition of the DEP, the reaction process is modeled by using rejection sampling. Thus, if there exists at least one infected individual, one should generate a random number $x$ in the $[0,1)$ interval for each susceptible individual and if $x\leq \mu_c$, it becomes an infected one. Simultaneously, one should generate a random number $x$ in the $[0,1)$ interval for each infected individual and if $x\leq \mu_r$, the infected individual becomes a susceptible one. The control parameter is the concentration $\rho$, and we randomly distribute a population of walkers
\begin{equation}
  N_w = N\rho,
\label{walker-pop}
\end{equation}
in the network at $t=0$ with half infected and half susceptible. However, the original definition does not have a finite threshold, and this means that any infection can survive in the Apollonian network if we allow unrestricted contacts between the individuals in the same node.

We introduce a modification to the original definition that transforms the $x\leq \mu_c$ and $x\leq \mu_r$ probabilities into taxes and the contamination and cure processes inside a node are simulated by the Gillespie Algorithm. Next, we enumerate the following rules which define the MDEP model applied to Apollonian networks \cite{tfaalves2020}:

\begin{enumerate}
\item \textbf{Initialization step:} At time $t=0$, a population $N_w$ of walkers, given as a function of the concentration in Eq.(\ref{walker-pop}), are randomly distributed in the nodes of an Apollonian network and half of the population is randomly chosen to be infected. The numbers of susceptible and infected individuals in each node $i$ are stored in the two arrays $\mathbf{A} \equiv \left\lbrace A(1), A(2), ..., A(N)\right\rbrace$, and $\mathbf{B} \equiv \left\lbrace B(1), B(2), ..., B(N)\right\rbrace$, respectively. Along the dynamics, we must count the number of visits $N_{\mathrm{r}}$ to the absorbing state. Thus, in the beginning of the simulation, we set $N_{\mathrm{r}}=0$;
\item \textbf{Evolution step:} One evolution step is divided in two stages, where all network nodes are updated simultaneously. The arrays $\mathbf{A}$ and $\mathbf{B}$ are updated at the end of each stage:
   \begin{itemize}
     \item \textbf{Diffusion stage:} One should generate a random uniform number $x$ in the $[0,1)$ interval for each susceptible individual in the node $i$ and if $x\leq D_{A}$ (where $D_{\textrm{A}} \in [0,1]$), a susceptible particle jumps from the node $i$ to a randomly chosen neighboring node $j$, in such a way that the arrays are updated as follows
     \begin{eqnarray}
        A(i) &=& A(i) - 1, \nonumber \\
        A(j) &=& A(j) + 1,
     \end{eqnarray}
and if a random number uniformly distributed on $[0,1)$ interval is less or equal to $D_{\textrm{B}}$ (where $D_{\textrm{B}} \in [0,1]$) the infected particle jumps from the node $i$ to a randomly chosen neighboring node $j$. Then, the infected populations are updated as follows
     \begin{eqnarray}
        B(i) &=& B(i) - 1, \nonumber \\
        B(j) &=& B(j) + 1.
     \end{eqnarray}
     \item \textbf{Reaction stage:} The time evolution of the populations in each node $i$ is stochastically simulated by using Gillespie algorithm in a time $t_\mathrm{max}$, exponentially distributed with mean $1/P(i)$, i.e.
\begin{equation}
t_\mathrm{max} = -\frac{1}{N_w(i)}\ln(1-x),
\label{exponentialdistrib_tmax}
\end{equation}
where $x$ is a random number in the interval $[0,1)$ and $N_w(i)=A(i)+B(i)$ is the population of the node $i$. The populations in each compartment are treated as reactants;
      \begin{enumerate}
         \item \textbf{Initialization}: The initial reactants $A(i,0)$ and $B(i,0)$ are set to the populations $A(i)$ and $B(i)$ after the diffusion stage, and the reaction time $t_q$ is set to zero;
         \item \textbf{Monte-Carlo step}: One random number in the interval $[0,1)$ is generated to select if a contamination or a spontaneous recover will take place, with probabilities proportional to its propensities. The contamination channel
            \begin{eqnarray}
            A(i,t_q+\Delta t_q) &=& A(i,t_q) - 1, \nonumber \\
            B(i,t_q+\Delta t_q) &=& B(i,t_q) + 1,
            \end{eqnarray}
have a propensity $\kappa(i,1)$ given by
            \begin{equation}
            \kappa(i,1) = \mu_c \frac{B(i,t_q)}{N_w(i,t_q)}A(i,t_q),
            \end{equation}
and the spontaneous recover channel
            \begin{eqnarray}
            A(i,t_q+\Delta t_q) &=& A(i,t_q) + 1, \nonumber \\
            B(i,t_q+\Delta t_q) &=& B(i,t_q) - 1,
            \end{eqnarray}
have a propensity $\kappa(i,2)$ given by
            \begin{equation}
            \kappa(i,2) = \mu_r B(i,t_q).
            \end{equation}
The reaction time $t_q$ is then updated by adding it with an exponentially distributed time interval $\Delta t_q$ with mean given by $1/(\kappa(i,1)+\kappa(i,2))$;
            \item \textbf{Iteration}: Step (b) is repeated until the reaction time $t_q$ exceeds $t_\mathrm{max}$ or if there is not any infected individual in the node $i$. When the reaction ends, $A(i)$ and $B(i)$ are updated with the values of $A(i,t_q)$ and $B(i,t_q)$;
      \end{enumerate}
   \end{itemize}
\item \textbf{Reactivation step:} The simulation time is then updated by a time unit. If there is no infected individual in the entire network, we increase $N_{\mathrm{r}}$ by one unit, and we randomly select one node of the network and turn all of its susceptible individuals to infected ones in order to continue the simulation \cite{macedo2018};
\item \textbf{DEP Iteration:} Steps 2 and 3 are repeated until the system reaches a stationary state.
\end{enumerate}

Our approach is a mixed one, that can be interpreted as a diffusion coupled agent model. The propagation is modeled with diffusion probabilities by rejection sampling and the reaction process are simulated by contacts in each node by using Gillespie algorithm where the number of contacts are a result of the competition between the two time scales: $t_\mathrm{max}$, and $1/(\kappa(i,1)+\kappa(i,2))$. This mixed approach is identical to an algorithm that simulates the chemical master equation by synchronizing the diffusion and reaction time scales, applied to the simulation of cell cultures\cite{Shimoni-2011}. At last, the mixed approach allows us to investigate the rich critical behavior of the MDEP model, which is dependent on the discrete time diffusion rates for each $D_{A}<D_{B}$, $D_{A}=D_{B}$, and $D_{A}>D_{B}$ regime, respectively.

To implement the MDEP model on the Apollonian network we build networks of size $N=1096, 3283, 9844$ and $29527$ corresponding to generations $n=7,8,9$ and $10$, respectively. In order to obtain the relevant observables, we used MC simulation and Finite Size-Scaling collapses. We considered only three cases of different diffusion rates given by $D_{A}=0.25$ and $D_{B}=0.75$, $D_{A}=0.50$ and $D_{B}=0.50$, and $D_{A}=0.75$ and $D_{B}=0.25$ to investigate if the critical exponent ratios should depend on them. For each network, we considered $10^5$ MC steps to let the system evolve to a stationary state and another $10^7$ MC steps to collect $10^7$ values of the observables written on.

\section{\label{sec:level3}Results and Discussions}

In this section, we focus the attention on our main MC simulation results. Thus, we present the results for the MDEP model on the Apollonian network. First, we employ a finite-size scaling analysis (FSS) to estimate the critical point, and to directly obtain the critical exponents $1/\nu$, $\beta/\nu$, and $\gamma/\nu$ for each $D_{A}<D_{B}$, $D_{A}=D_{B}$, and $D_{A}>D_{B}$ regime, respectively. In order to investigate the MDEP critical behavior, one can obtain a time series of the following observables at the stationary state
\begin{eqnarray}
\rho_B &=& \frac{1}{N_w}\sum_i^N B(i) \nonumber \\
\rho'_B &=& \frac{1}{N}\sum_i^N \left(1-\delta_{B(i),0}\right),
\label{observables}
\end{eqnarray}
which are the infection density and the fraction of active nodes, i.e., nodes with at least one infected individual. The following averages from the time series of the infection density $\rho_{B}$ on the stationary state can be obtained as functions of the density of individuals $\rho$:
\begin{eqnarray}
P &=& \left< \rho_{B} \right>, \nonumber \\
U &=&\frac{\left< \rho^{2}_{B} \right>\left< \rho^{3}_{B} \right>  - \left< \rho^{ }_{B} \right>\left< \rho^{2}_{B} \right>^{2}}
          {\left< \rho^{ }_{B} \right>\left< \rho^{4}_{B} \right>  - \left< \rho^{ }_{B} \right>\left< \rho^{2}_{B} \right>^{2}}, \nonumber \\
\Delta &=& N_w \left( \left< \rho^{2}_{B} \right> - \left< \rho^{ }_{B} \right>^{2} \right).
\label{dep-averages}
\end{eqnarray}
Here, $P$ is the order parameter, $U$ is the $5$-order cumulant ratio for directed percolation, and $\Delta$ is the order parameter fluctuation. The $5$-order cumulant ratio is finite at the absorbing phase, and crosses on distinct network size data at the collective critical threshold\cite{lubeck2002, jansen2007, henkel2008}. Analogous averages $P_\mathrm{node}$, $U_\mathrm{node}$, and $\Delta_\mathrm{node}$ can be constructed from the fraction of active nodes, which have the same critical behavior of $P$, $U$, and $\Delta$, respectively.

We conjecture that the averages shown on Eq.(\ref{dep-averages}) obey the following FSS relations
\begin{eqnarray}
\rho_{B} &\approx& N^{-\beta/\nu} f_{\rho_{B}}\left[N^{1/\nu}\left(\rho-\rho_{c}\right)\right], \nonumber \\
U &\approx& f_{U}\left[N^{1/\nu}\left(\rho-\rho_{c}\right)\right], \nonumber \\
\Delta &\approx& N^{\gamma/\nu} f_{\Delta}\left[N^{1/\nu}\left(\rho-\rho_{c}\right)\right],
\label{dep-fss}
\end{eqnarray}
close to the critical threshold $\rho_c$, where $1/\nu$, $\beta/\nu$, and $\gamma/\nu$ are the critical exponent ratios, and $f_{\rho_{B},U,\Delta}$ are, respectively, the FSS functions.

\begin{figure}
\includegraphics[scale=0.16]{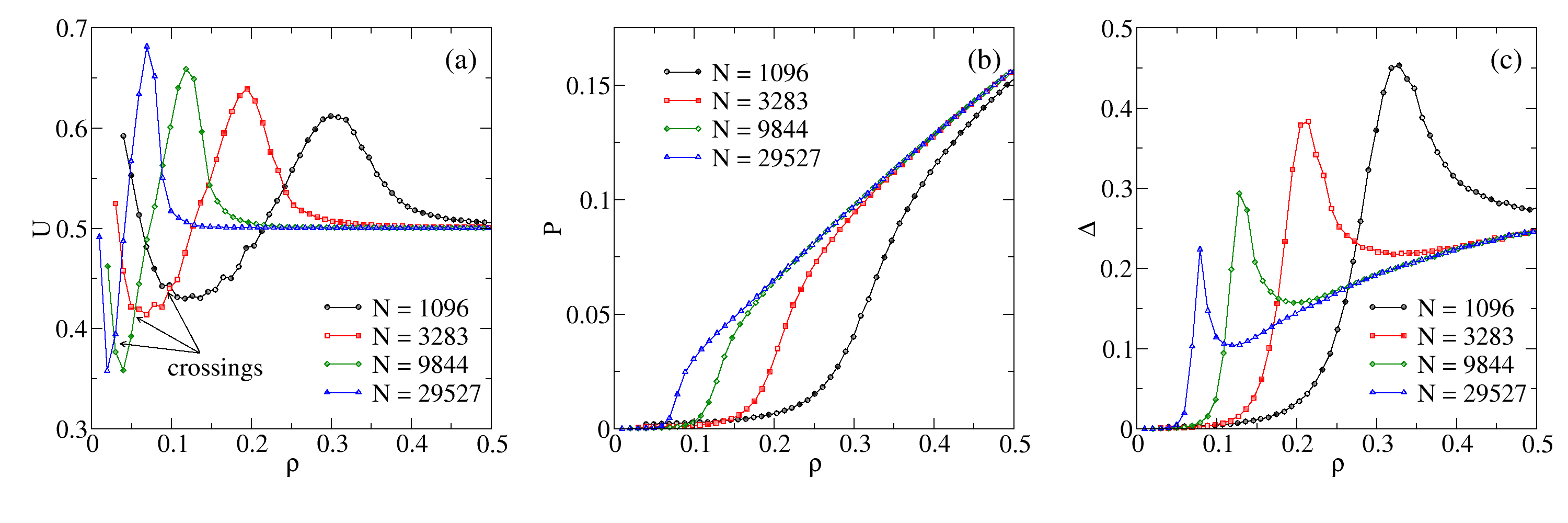}
\caption{(Color online). DEP on the Apollonian network for generations 7 to 10 with parameters $\mu_c=\mu_r=D_A=D_B=0.5$. Note that the crossings of the cumulant for increasing network sizes become closer to zero density, which is consistent with an active phase for any non-zero density in the infinite network size limit. In addition, the order parameter and its fluctuations are compatible with monotonic increasing functions in the infinite network size limit.}
\label{fig:dep-original}
\end{figure}

Now, we turn to the simulation results. First, we show an example of DEP dynamics on Apollonian networks in Fig.(\ref{fig:dep-original}) for $\mu_c=\mu_r=D_A=D_B=0.5$. Note that the crossings for increasing network sizes are closer to zero concentration which allows concluding that the system is active for any finite concentration in the infinite network limit. In addition, the order parameter and its fluctuations are compatible with monotonic increasing functions in the infinite lattice size limit. In particular, the inflection points of the order parameter curves, and the peaks of the order parameter fluctuations goes closer to zero when increasing the network size. 

\begin{figure}
\includegraphics[scale=0.16]{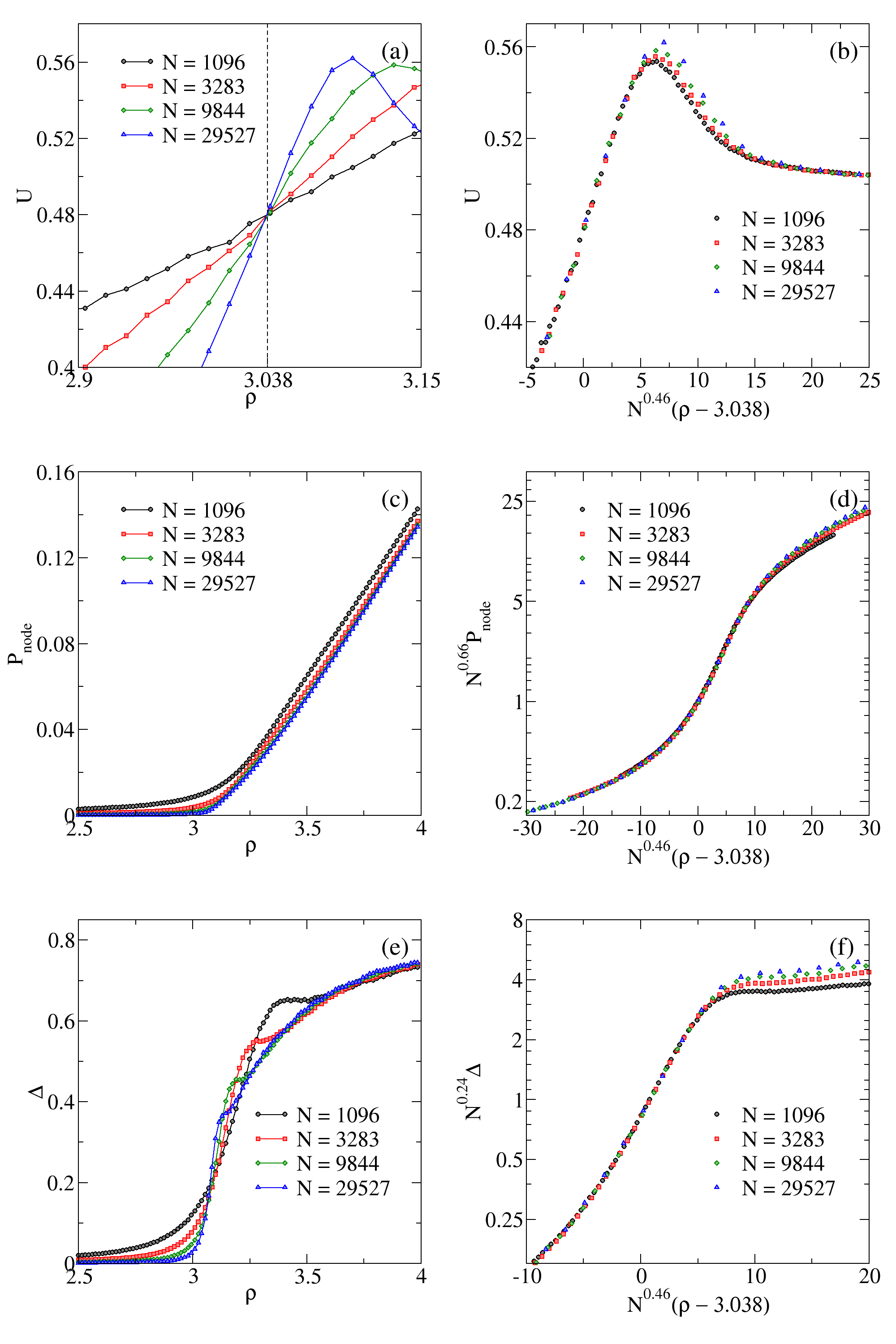}
\caption{(Color online). Critical behavior of the MDEP model for the $D_{A}=0.25$ and $D_{B}=0.75$ rates. Panels (a), (c), and (e) show the Binder's $5$-order cumulant U, the order parameter $P_\mathrm{node}$ and the order parameter fluctuations $\Delta$, both as function of the concentration $\rho$ for several network sizes. Panel (c) shows a sigmoidal shape for the order parameter and a marked inflection point characterizing the critical behavior and panel (e) is compatible with a finite jump for the order parameter fluctuations. Panels (b), (d), and (f) show the best data collapse by using the FSS analysis, from which we drew up the critical exponents $1/\nu=0.46(2)$, $\beta/\nu=0.66(2)$, and $\gamma/\nu=-0.24(2)$. The critical threshold $\rho_{c}=3.038(1)$ was obtained from Binder's $5$-order cumulant.}
\label{fig:025075}
\end{figure}

\begin{figure}
\includegraphics[scale=0.16]{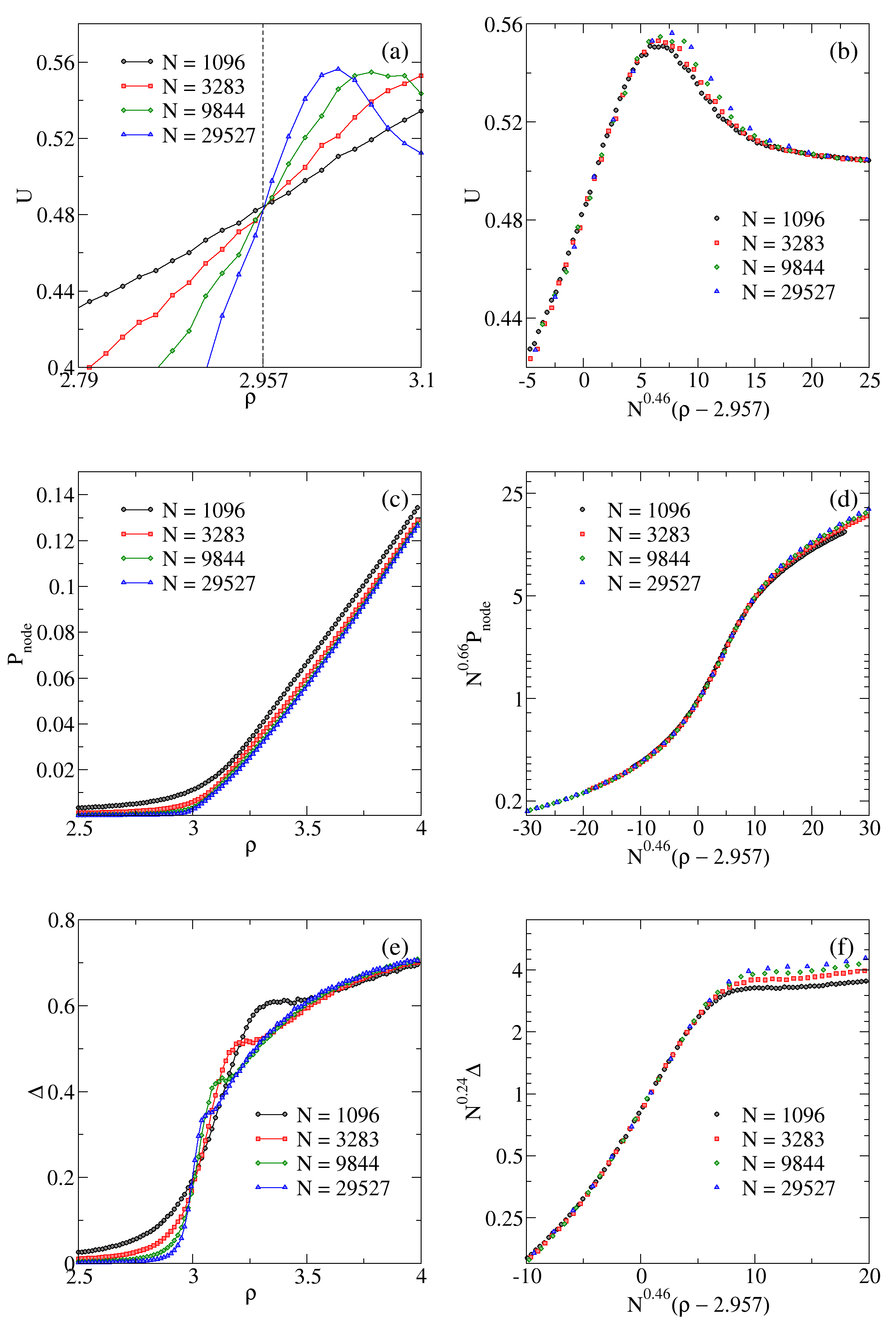}
\caption{(Color online). The same of Fig.(\ref{fig:025075}) for $D_{A}=0.50$ and $D_{B}=0.50$ rates. The critical threshold for this parameter set is $\rho_{c}=2.957(1)$.}
\label{fig:050050} 
\end{figure}

\begin{figure}
\includegraphics[scale=0.16]{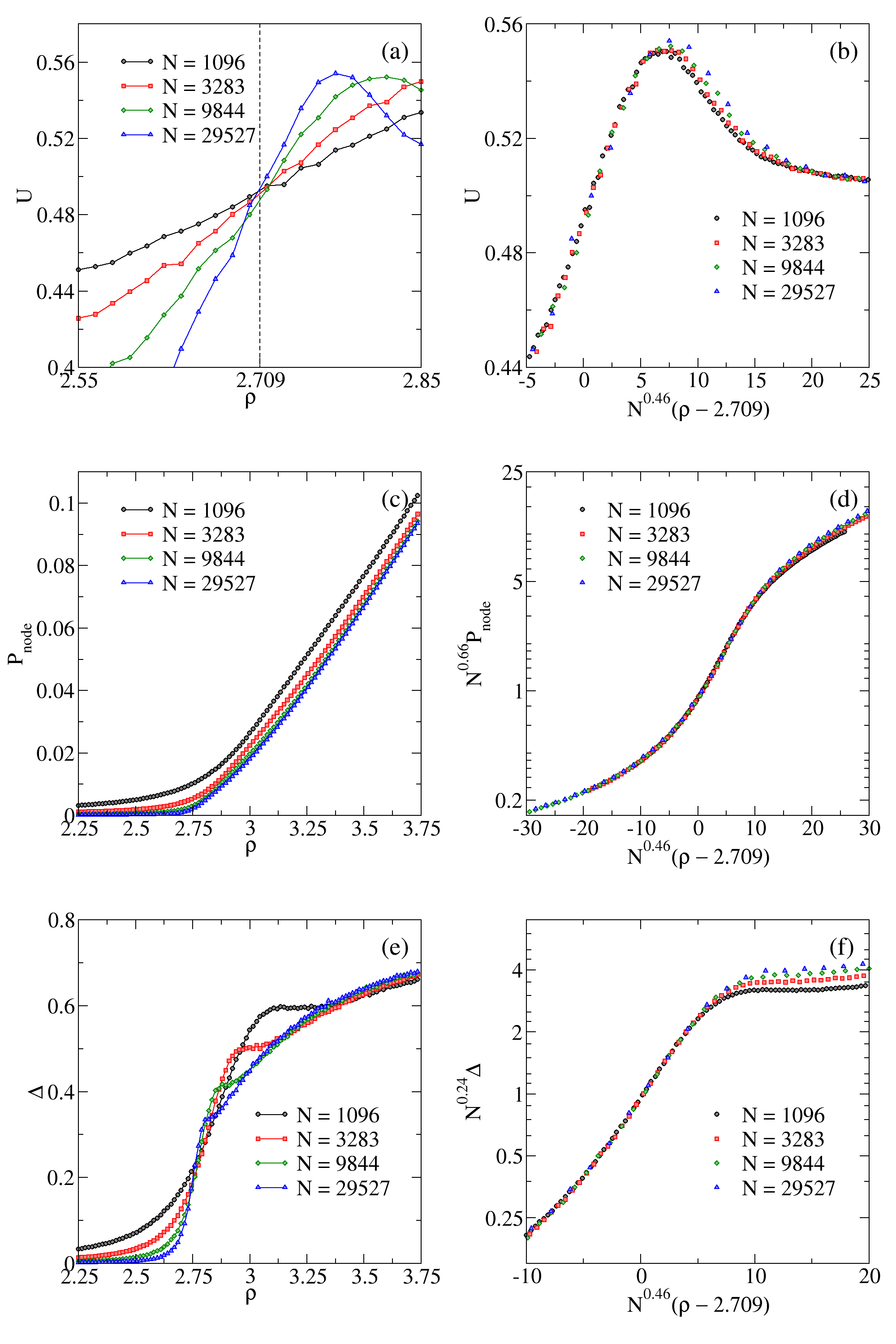}
\caption{(Color online). The same of Fig.(\ref{fig:025075}) for $D_{A}=0.75$ and $D_{B}=0.25$ rates. The critical threshold  for this parameter set is $\rho_{c}=2.709(1)$.}
\label{fig:075025}
\end{figure}

Continuing the discussion, we show results for the MDEP in Figs.(\ref{fig:025075}), (\ref{fig:050050}), and (\ref{fig:075025}) for $D_{A}>D_{B}$, $D_{A}=D_{B}$, and $D_{A}<D_{B}$ regimes, respectively. In all cases, we considered contamination and recovering taxes $\mu_c=2$, and $\mu_r=1$, respectively. In panel (a) of Figs.(\ref{fig:025075}), (\ref{fig:050050}), and (\ref{fig:075025}), we show the $5$-order cumulant curves and by the cumulant crossings, we estimate the collective thresholds $\rho_{c}=3.038(1)$, $\rho_{c}=2.957(1)$, and $\rho_{c}=2.709(3)$, respectively. In panels (c) of Figs.(\ref{fig:025075}), (\ref{fig:050050}), and (\ref{fig:075025}), we show the average of the fraction of active nodes where we can see the typical sigmoidal shape where the curve inflection separates the absorbing and active phases. Particularly interesting is the behavior of the order parameter fluctuations, which is compatible with a finite jump, instead of a typical peak that increases with the network size.

Regarding the critical behavior results, in panels (b), (d), and (f) of Figs.(\ref{fig:025075}), (\ref{fig:050050}), and (\ref{fig:075025}), we show the best data collapses of the cumulant, the active node fraction average and the order parameter fluctuations by the use of each previously critical thresholds $\rho_{c}$ computed and estimated the following critical exponent ratios: $\beta/\nu=0.66(2)$, $1/\nu=0.46(2)$, and $\gamma/\nu=-0.24(2)$ in all the three $D_{A}>D_{B}$, $D_{A}=D_{B}$, and $D_{A}<D_{B}$ regimes. The $\gamma$ exponent is negative, corresponding to a finite jump, instead of a divergence with the network size.

Additionally, reporting the literature about simulation studies on low dimensional regular lattices for DEP model, all investigations exhibit a continuous phase transition to an absorbing state with well defined critical properties, characterizing distinct universality classes. This is in contrast to van Wijland et al. to $D_{A}>D_{B}$ regime\cite{fulco2001,maia2007,amfilho2010} which predict a discontinuous phase transition. Regarding complex networks, CP model on the Apollonian network presented a continuous phase transition with well defined critical point and exponents \cite{lfsilva2013}. They concluded that the Apollonian network topology, with $2 < \gamma \leq 3$, cannot affect the overall epidemic spreading, and that the CP model on this network presents a phase transition with critical exponents $\beta/\nu=0.54(2)$ and $1/\nu=0.51(2)$, closer to the MF exponent for the $\gamma > 3$, corroborating with the universality class of the MF theory of regular lattice.

Recently, the MDEP model was implemented on Barabasi-Albert network \cite{tfaalves2020}. Here, was analyzed a MDEP furnishing a continuous phase transition obeying the MF critical exponents $\beta=1$, $\nu=0.5$, and $\gamma=0$, but with logarithmic corrections to order parameter and its fluctuations to all $D_{A}<D_{B}$, $D_{A}=D_{B}$, and $D_{A}>D_{B}$ studied regimes. In this context, although the MDEP model on the Apollonian network presents continuous phase transition to all diffusion regimes, it does not agree with the literature, exhibiting a new set of critical exponents which are different from MF theory, both characterizing the different diffusive regimes investigated in this work. 

\section{\label{sec:level4}Summary and Conclusions}

In this work, we study the DEP model on the Apollonian network by implementing a new approach, based on Gillespie algorithm which introduces a modification in the reaction process, limiting the contacts among individuals into a same site and, consequently, introducing a phase transition that is not present if the contacts are unrestricted. In this model, each individual may diffuse on the network with constant diffusion rates $D_{A}$ and $D_{B}$. Here, for the first time, the DEP model displays a phase transition on the Apollonian network.

Regarding the epidemic spreading, the model is compatible with the fact that restricting the contacts between individuals can allow for epidemic control in a way that the system can present a critical threshold between the absorbing phase and the active phase by increasing the populational concentration. In this way, the model favors a social distancing behavior in order to allow epidemic control. In the converse, unrestricted contacts lead to epidemic survival for any non-zero concentration in the infinite network limit.

In contrast to van Wijland et al. to $D_{A}>D_{B}$ regime, the MDEP model presents a continuous phase transition to an absorbing state being characterized by the same set critical exponents, for every investigated diffusive regime. The critical exponent ratios are given by $\beta/\nu=0.66(2)$, $1/\nu=0.46(2)$, and $\gamma/\nu=-0.24(2)$. This set of critical exponents does not belong to MF universality class in both regular lattices and complex networks.

\begin{acknowledgments}

We would like to thank CAPES (Coordena\c{c}\~{a}o de Aperfei\c{c}oamento de Pessoal de N\'{\i}vel Superior), CNPq (Conselho Nacional de Desenvolvimento Cient\'{\i}fico e tecnol\'{o}gico), FUNCAP (Funda\c{c}\~{a}o Cearense de Apoio ao Desenvolvimento Cient\'{\i}fico e Tecnol\'{o}gico), and FAPEPI (Funda\c{c}\~{a}o de Amparo a Pesquisa do Estado do Piau\'{\i}) for the financial support. We acknowledge the use of Dietrich Stauffer Computational Physics Lab, Teresina, Brazil, and Laborat\'{o}rio de F\'{\i}sica Te\'{o}rica e Modelagem Computacional - LFTMC, Teresina, Brazil, where the numerical simulations were performed.

\end{acknowledgments}

\bibliography{draft_depan}

\providecommand{\noopsort}[1]{}\providecommand{\singleletter}[1]{#1}%
\begin{thebibliography}{44}
\expandafter\ifx\csname natexlab\endcsname\relax\def\natexlab#1{#1}\fi
\expandafter\ifx\csname bibnamefont\endcsname\relax
  \def\bibnamefont#1{#1}\fi
\expandafter\ifx\csname bibfnamefont\endcsname\relax
  \def\bibfnamefont#1{#1}\fi
\expandafter\ifx\csname citenamefont\endcsname\relax
  \def\citenamefont#1{#1}\fi
\expandafter\ifx\csname url\endcsname\relax
  \def\url#1{\texttt{#1}}\fi
\expandafter\ifx\csname urlprefix\endcsname\relax\def\urlprefix{URL }\fi
\providecommand{\bibinfo}[2]{#2}
\providecommand{\eprint}[2][]{\url{#2}}

\bibitem[{\citenamefont{Zaher et~al.}(2020)\citenamefont{Zaher, Mostafa, and
  Altaher}}]{zaher2020}
\bibinfo{author}{\bibfnamefont{N.~H.} \bibnamefont{Zaher}},
  \bibinfo{author}{\bibfnamefont{M.~I.} \bibnamefont{Mostafa}},
  \bibnamefont{and} \bibinfo{author}{\bibfnamefont{A.~Y.}
  \bibnamefont{Altaher}}, \bibinfo{journal}{Acta Pharm.}
  \textbf{\bibinfo{volume}{70}}, \bibinfo{pages}{145} (\bibinfo{year}{2020}).

\bibitem[{\citenamefont{Sun et~al.}(2020)\citenamefont{Sun, Wang, Huang, Wang,
  Cao, Zhang, Zheng, and Lu}}]{sun2020}
\bibinfo{author}{\bibfnamefont{W.}~\bibnamefont{Sun}},
  \bibinfo{author}{\bibfnamefont{L.}~\bibnamefont{Wang}},
  \bibinfo{author}{\bibfnamefont{H.}~\bibnamefont{Huang}},
  \bibinfo{author}{\bibfnamefont{W.}~\bibnamefont{Wang}},
  \bibinfo{author}{\bibfnamefont{L.}~\bibnamefont{Cao}},
  \bibinfo{author}{\bibfnamefont{J.}~\bibnamefont{Zhang}},
  \bibinfo{author}{\bibfnamefont{M.}~\bibnamefont{Zheng}}, \bibnamefont{and}
  \bibinfo{author}{\bibfnamefont{H.}~\bibnamefont{Lu}}, \bibinfo{journal}{Virus
  Research} \textbf{\bibinfo{volume}{278}}, \bibinfo{pages}{197869}
  (\bibinfo{year}{2020}).

\bibitem[{\citenamefont{Lu et~al.}(2020)\citenamefont{Lu, Chen, Shao, Dong,
  Liu, and Li}}]{lu2020}
\bibinfo{author}{\bibfnamefont{G.}~\bibnamefont{Lu}},
  \bibinfo{author}{\bibfnamefont{R.}~\bibnamefont{Chen}},
  \bibinfo{author}{\bibfnamefont{R.}~\bibnamefont{Shao}},
  \bibinfo{author}{\bibfnamefont{N.}~\bibnamefont{Dong}},
  \bibinfo{author}{\bibfnamefont{W.}~\bibnamefont{Liu}}, \bibnamefont{and}
  \bibinfo{author}{\bibfnamefont{S.}~\bibnamefont{Li}},
  \bibinfo{journal}{Journal of Infection} \textbf{\bibinfo{volume}{80}},
  \bibinfo{pages}{350} (\bibinfo{year}{2020}).

\bibitem[{\citenamefont{Zhang and Liu}(2020)}]{zhang2020}
\bibinfo{author}{\bibfnamefont{L.}~\bibnamefont{Zhang}} \bibnamefont{and}
  \bibinfo{author}{\bibfnamefont{Y.}~\bibnamefont{Liu}},
  \bibinfo{journal}{Journal of Medical Viruses} \textbf{\bibinfo{volume}{92}},
  \bibinfo{pages}{479} (\bibinfo{year}{2020}).

\bibitem[{\citenamefont{Bartlett}(1956)}]{bartlett1956}
\bibinfo{author}{\bibfnamefont{M.~S.} \bibnamefont{Bartlett}},
  \bibinfo{journal}{Proc. Third Berkeley Symp. Math. Stat. Prob.}
  \textbf{\bibinfo{volume}{4}}, \bibinfo{pages}{81} (\bibinfo{year}{1956}).

\bibitem[{\citenamefont{Castellano and Pastor-Satorras}(2020)}]{castellano2020}
\bibinfo{author}{\bibfnamefont{C.}~\bibnamefont{Castellano}} \bibnamefont{and}
  \bibinfo{author}{\bibfnamefont{R.}~\bibnamefont{Pastor-Satorras}},
  \bibinfo{journal}{Phys. Rev. X} \textbf{\bibinfo{volume}{10}},
  \bibinfo{pages}{011070} (\bibinfo{year}{2020}).

\bibitem[{\citenamefont{Mota et~al.}(2018)\citenamefont{Mota, Alves,
  Macedo-Filho, and Alves}}]{mota2018}
\bibinfo{author}{\bibfnamefont{M.~P.~S.} \bibnamefont{Mota}},
  \bibinfo{author}{\bibfnamefont{G.~A.} \bibnamefont{Alves}},
  \bibinfo{author}{\bibfnamefont{A.}~\bibnamefont{Macedo-Filho}},
  \bibnamefont{and} \bibinfo{author}{\bibfnamefont{T.~F.~A.}
  \bibnamefont{Alves}}, \bibinfo{journal}{Phy. A}
  \textbf{\bibinfo{volume}{510}}, \bibinfo{pages}{577} (\bibinfo{year}{2018}).

\bibitem[{\citenamefont{Macedo-Filho et~al.}(2018)\citenamefont{Macedo-Filho,
  Alves, Filho, and Alves}}]{macedo2018}
\bibinfo{author}{\bibfnamefont{A.}~\bibnamefont{Macedo-Filho}},
  \bibinfo{author}{\bibfnamefont{G.~A.} \bibnamefont{Alves}},
  \bibinfo{author}{\bibfnamefont{R.~N.~C.} \bibnamefont{Filho}},
  \bibnamefont{and} \bibinfo{author}{\bibfnamefont{T.~F.~A.}
  \bibnamefont{Alves}}, \bibinfo{journal}{J. of Stat. Mech.}
  \textbf{\bibinfo{volume}{2018}}, \bibinfo{pages}{043208}
  (\bibinfo{year}{2018}).

\bibitem[{\citenamefont{Bailey}(1964)}]{bailey1957}
\bibinfo{author}{\bibfnamefont{N.~T.~J.} \bibnamefont{Bailey}},
  \emph{\bibinfo{title}{The Elements of Stochastic Processes with Applications
  to the Natural Sciences}} (\bibinfo{publisher}{John Wiley and Sons},
  \bibinfo{address}{New York}, \bibinfo{year}{1964}).

\bibitem[{\citenamefont{Santos et~al.}(2020)\citenamefont{Santos, Alves, Alves,
  Macedo-Filho, and Ferreira}}]{santos2020}
\bibinfo{author}{\bibfnamefont{G.~B.} \bibnamefont{Santos}},
  \bibinfo{author}{\bibfnamefont{G.~A.} \bibnamefont{Alves}},
  \bibinfo{author}{\bibfnamefont{T.~F.~A.} \bibnamefont{Alves}},
  \bibinfo{author}{\bibfnamefont{A.}~\bibnamefont{Macedo-Filho}},
  \bibnamefont{and} \bibinfo{author}{\bibfnamefont{R.~S.}
  \bibnamefont{Ferreira}}, \bibinfo{journal}{Phys. Lett. A}
  \textbf{\bibinfo{volume}{384}}, \bibinfo{pages}{126063}
  (\bibinfo{year}{2020}).

\bibitem[{\citenamefont{Alencar et~al.}(2020)\citenamefont{Alencar, Alves,
  Alves, Macedo-Filho, and Ferreira}}]{alencar2020}
\bibinfo{author}{\bibfnamefont{D.~S.~M.} \bibnamefont{Alencar}},
  \bibinfo{author}{\bibfnamefont{T.~F.~A.} \bibnamefont{Alves}},
  \bibinfo{author}{\bibfnamefont{G.}~\bibnamefont{Alves}},
  \bibinfo{author}{\bibfnamefont{A.}~\bibnamefont{Macedo-Filho}},
  \bibnamefont{and} \bibinfo{author}{\bibfnamefont{R.}~\bibnamefont{Ferreira}},
  \bibinfo{journal}{Phys. A} \textbf{\bibinfo{volume}{541}},
  \bibinfo{pages}{122800} (\bibinfo{year}{2020}).

\bibitem[{\citenamefont{Harris}(1974)}]{harris1974}
\bibinfo{author}{\bibfnamefont{T.}~\bibnamefont{Harris}},
  \bibinfo{journal}{Ann. Probab.} \textbf{\bibinfo{volume}{2}},
  \bibinfo{pages}{969} (\bibinfo{year}{1974}).

\bibitem[{\citenamefont{da~Silva et~al.}(2013)\citenamefont{da~Silva, Filho,
  Soares, Macedo-Filho, Fulco, and Albuquerque}}]{lfsilva2013}
\bibinfo{author}{\bibfnamefont{L.~F.} \bibnamefont{da~Silva}},
  \bibinfo{author}{\bibfnamefont{R.~N.~C.} \bibnamefont{Filho}},
  \bibinfo{author}{\bibfnamefont{D.~J.~B.} \bibnamefont{Soares}},
  \bibinfo{author}{\bibfnamefont{A.}~\bibnamefont{Macedo-Filho}},
  \bibinfo{author}{\bibfnamefont{U.}~\bibnamefont{Fulco}}, \bibnamefont{and}
  \bibinfo{author}{\bibfnamefont{E.~L.} \bibnamefont{Albuquerque}},
  \bibinfo{journal}{Phys. A} \textbf{\bibinfo{volume}{392}},
  \bibinfo{pages}{1532} (\bibinfo{year}{2013}).

\bibitem[{\citenamefont{Marro and Dickman}(1999)}]{marro1999}
\bibinfo{author}{\bibfnamefont{J.}~\bibnamefont{Marro}} \bibnamefont{and}
  \bibinfo{author}{\bibfnamefont{R.}~\bibnamefont{Dickman}},
  \emph{\bibinfo{title}{Nonequilibrium Phase Transitions in Lattice Models}}
  (\bibinfo{publisher}{Cambridge University Press},
  \bibinfo{address}{Cambridge}, \bibinfo{year}{1999}).

\bibitem[{\citenamefont{Almeida et~al.}(2016)\citenamefont{Almeida,
  Macedo-Filho, Mendes, da~Silva, Albuquerque, and Fulco}}]{almeida2016}
\bibinfo{author}{\bibfnamefont{M.~L.} \bibnamefont{Almeida}},
  \bibinfo{author}{\bibfnamefont{A.}~\bibnamefont{Macedo-Filho}},
  \bibinfo{author}{\bibfnamefont{G.~A.} \bibnamefont{Mendes}},
  \bibinfo{author}{\bibfnamefont{L.~R.} \bibnamefont{da~Silva}},
  \bibinfo{author}{\bibfnamefont{E.~L.} \bibnamefont{Albuquerque}},
  \bibnamefont{and} \bibinfo{author}{\bibfnamefont{U.~L.} \bibnamefont{Fulco}},
  \bibinfo{journal}{J. of Stat. Mech.} \textbf{\bibinfo{volume}{2016}},
  \bibinfo{pages}{043202} (\bibinfo{year}{2016}).

\bibitem[{\citenamefont{Kree et~al.}(1989)\citenamefont{Kree, Schaub, and
  Schmittmann}}]{kree1989}
\bibinfo{author}{\bibfnamefont{R.}~\bibnamefont{Kree}},
  \bibinfo{author}{\bibfnamefont{B.}~\bibnamefont{Schaub}}, \bibnamefont{and}
  \bibinfo{author}{\bibfnamefont{B.}~\bibnamefont{Schmittmann}},
  \bibinfo{journal}{Phys. Rev. A} \textbf{\bibinfo{volume}{39}},
  \bibinfo{pages}{214} (\bibinfo{year}{1989}).

\bibitem[{\citenamefont{van Wijland et~al.}(1998)\citenamefont{van Wijland,
  Oerding, and Hilhorst}}]{wijland1998}
\bibinfo{author}{\bibfnamefont{F.}~\bibnamefont{van Wijland}},
  \bibinfo{author}{\bibfnamefont{K.}~\bibnamefont{Oerding}}, \bibnamefont{and}
  \bibinfo{author}{\bibfnamefont{H.~J.} \bibnamefont{Hilhorst}},
  \bibinfo{journal}{Phys. Rev. A} \textbf{\bibinfo{volume}{251}},
  \bibinfo{pages}{179} (\bibinfo{year}{1998}).

\bibitem[{\citenamefont{Fulco et~al.}(2001)\citenamefont{Fulco, Messias, and
  Lyra}}]{fulco2001}
\bibinfo{author}{\bibfnamefont{U.~L.} \bibnamefont{Fulco}},
  \bibinfo{author}{\bibfnamefont{D.~N.} \bibnamefont{Messias}},
  \bibnamefont{and} \bibinfo{author}{\bibfnamefont{M.~L.} \bibnamefont{Lyra}},
  \bibinfo{journal}{Phys. Rev. E} \textbf{\bibinfo{volume}{63}},
  \bibinfo{pages}{066118} (\bibinfo{year}{2001}).

\bibitem[{\citenamefont{Maia and Dickman}(2007)}]{maia2007}
\bibinfo{author}{\bibfnamefont{D.~S.} \bibnamefont{Maia}} \bibnamefont{and}
  \bibinfo{author}{\bibfnamefont{R.}~\bibnamefont{Dickman}},
  \bibinfo{journal}{J. Phys.: Condens. Matter} \textbf{\bibinfo{volume}{19}},
  \bibinfo{pages}{065143} (\bibinfo{year}{2007}).

\bibitem[{\citenamefont{Filho et~al.}(2010)\citenamefont{Filho, Corso, Fulco,
  and LYRA}}]{amfilho2010}
\bibinfo{author}{\bibfnamefont{A.~M.} \bibnamefont{Filho}},
  \bibinfo{author}{\bibfnamefont{G.}~\bibnamefont{Corso}},
  \bibinfo{author}{\bibfnamefont{U.~L.} \bibnamefont{Fulco}}, \bibnamefont{and}
  \bibinfo{author}{\bibfnamefont{M.~L.} \bibnamefont{LYRA}},
  \bibinfo{journal}{J. of Stat. Mech.} \textbf{\bibinfo{volume}{2010}}
  (\bibinfo{year}{2010}).

\bibitem[{\citenamefont{Hinrichsen}(2000)}]{hinrichsen2000}
\bibinfo{author}{\bibfnamefont{H.}~\bibnamefont{Hinrichsen}},
  \bibinfo{journal}{Advances in Physics} \textbf{\bibinfo{volume}{49}},
  \bibinfo{pages}{815} (\bibinfo{year}{2000}).

\bibitem[{\citenamefont{Jr. et~al.}(2005)\citenamefont{Jr., Herrmann, Andrade,
  and da~Silva}}]{jsandrade2005}
\bibinfo{author}{\bibfnamefont{J.~S.~A.} \bibnamefont{Jr.}},
  \bibinfo{author}{\bibfnamefont{H.~J.} \bibnamefont{Herrmann}},
  \bibinfo{author}{\bibfnamefont{R.~F.~S.} \bibnamefont{Andrade}},
  \bibnamefont{and} \bibinfo{author}{\bibfnamefont{L.~R.}
  \bibnamefont{da~Silva}}, \bibinfo{journal}{Phys. Rev. Lett.}
  \textbf{\bibinfo{volume}{94}}, \bibinfo{pages}{018702}
  (\bibinfo{year}{2005}).

\bibitem[{\citenamefont{Ferreira
  et~al.}(2011{\natexlab{a}})\citenamefont{Ferreira, Ferreira, and
  Pastor-Satorras}}]{scferreira2011}
\bibinfo{author}{\bibfnamefont{S.~C.} \bibnamefont{Ferreira}},
  \bibinfo{author}{\bibfnamefont{R.~S.} \bibnamefont{Ferreira}},
  \bibnamefont{and}
  \bibinfo{author}{\bibfnamefont{R.}~\bibnamefont{Pastor-Satorras}},
  \bibinfo{journal}{Phys. Rev. E} \textbf{\bibinfo{volume}{83}},
  \bibinfo{pages}{066113} (\bibinfo{year}{2011}{\natexlab{a}}).

\bibitem[{\citenamefont{Noh and Park}(2009)}]{jdnoh2009}
\bibinfo{author}{\bibfnamefont{J.~D.} \bibnamefont{Noh}} \bibnamefont{and}
  \bibinfo{author}{\bibfnamefont{H.}~\bibnamefont{Park}},
  \bibinfo{journal}{Phys. Rev. E} \textbf{\bibinfo{volume}{79}},
  \bibinfo{pages}{056115} (\bibinfo{year}{2009}).

\bibitem[{\citenamefont{Ferreira
  et~al.}(2011{\natexlab{b}})\citenamefont{Ferreira, Ferreira, Castellano, and
  Pastor-Satorras}}]{rsferreira2011}
\bibinfo{author}{\bibfnamefont{S.}~\bibnamefont{Ferreira}},
  \bibinfo{author}{\bibfnamefont{R.}~\bibnamefont{Ferreira}},
  \bibinfo{author}{\bibfnamefont{C.}~\bibnamefont{Castellano}},
  \bibnamefont{and}
  \bibinfo{author}{\bibfnamefont{R.}~\bibnamefont{Pastor-Satorras}},
  \bibinfo{journal}{Phys. Rev. E} \textbf{\bibinfo{volume}{84}},
  \bibinfo{pages}{066102} (\bibinfo{year}{2011}{\natexlab{b}}).

\bibitem[{\citenamefont{Cohen and Havlin}(2010)}]{cohen2010}
\bibinfo{author}{\bibfnamefont{R.}~\bibnamefont{Cohen}} \bibnamefont{and}
  \bibinfo{author}{\bibfnamefont{S.}~\bibnamefont{Havlin}},
  \emph{\bibinfo{title}{Complex Networks: Structure, Robustness and Function}}
  (\bibinfo{publisher}{Cambridge University Press},
  \bibinfo{address}{Cambridge}, \bibinfo{year}{2010}).

\bibitem[{\citenamefont{Barab\'asi and P\'osfai}(2016)}]{barabasi2016}
\bibinfo{author}{\bibfnamefont{A.-L.} \bibnamefont{Barab\'asi}}
  \bibnamefont{and} \bibinfo{author}{\bibfnamefont{M.}~\bibnamefont{P\'osfai}},
  \emph{\bibinfo{title}{Network science}} (\bibinfo{publisher}{Cambridge
  University Press}, \bibinfo{address}{Cambridge}, \bibinfo{year}{2016}).

\bibitem[{\citenamefont{Liljeros et~al.}(2001)\citenamefont{Liljeros, Edling,
  Amaral, Stanley, and {\AA}berg}}]{lijeros2001}
\bibinfo{author}{\bibfnamefont{F.}~\bibnamefont{Liljeros}},
  \bibinfo{author}{\bibfnamefont{C.~R.} \bibnamefont{Edling}},
  \bibinfo{author}{\bibfnamefont{L.~A.~N.} \bibnamefont{Amaral}},
  \bibinfo{author}{\bibfnamefont{H.~E.} \bibnamefont{Stanley}},
  \bibnamefont{and}
  \bibinfo{author}{\bibfnamefont{Y.}~\bibnamefont{{\AA}berg}},
  \bibinfo{journal}{Nature} \textbf{\bibinfo{volume}{411}},
  \bibinfo{pages}{907} (\bibinfo{year}{2001}).

\bibitem[{\citenamefont{Mendes and da~Silva}(2009)}]{mendes2009}
\bibinfo{author}{\bibfnamefont{G.~A.} \bibnamefont{Mendes}} \bibnamefont{and}
  \bibinfo{author}{\bibfnamefont{L.~R.} \bibnamefont{da~Silva}},
  \bibinfo{journal}{Brazilian Journal of Physics}
  \textbf{\bibinfo{volume}{39}}, \bibinfo{pages}{423} (\bibinfo{year}{2009}).

\bibitem[{\citenamefont{Almeida et~al.}(2013)\citenamefont{Almeida, Mendes,
  Viswanatan, and da~Silva}}]{almeida2013}
\bibinfo{author}{\bibfnamefont{M.~L.} \bibnamefont{Almeida}},
  \bibinfo{author}{\bibfnamefont{G.~A.} \bibnamefont{Mendes}},
  \bibinfo{author}{\bibfnamefont{G.~M.} \bibnamefont{Viswanatan}},
  \bibnamefont{and} \bibinfo{author}{\bibfnamefont{L.~R.}
  \bibnamefont{da~Silva}}, \bibinfo{journal}{European Physical Journal B}
  \textbf{\bibinfo{volume}{86}}, \bibinfo{pages}{38} (\bibinfo{year}{2013}).

\bibitem[{\citenamefont{de~Solla~Price}(1965)}]{price1965}
\bibinfo{author}{\bibfnamefont{D.~J.} \bibnamefont{de~Solla~Price}},
  \bibinfo{journal}{Science} \textbf{\bibinfo{volume}{149}},
  \bibinfo{pages}{510} (\bibinfo{year}{1965}).

\bibitem[{\citenamefont{Gillespie and Petzold}(2013)}]{gillespie2013}
\bibinfo{author}{\bibfnamefont{D.~T.} \bibnamefont{Gillespie}}
  \bibnamefont{and} \bibinfo{author}{\bibfnamefont{L.~R.}
  \bibnamefont{Petzold}}, \bibinfo{journal}{J. Chem. Phys.}
  \textbf{\bibinfo{volume}{131}} (\bibinfo{year}{2013}).

\bibitem[{\citenamefont{Gillespie}(1976)}]{gillespie1976}
\bibinfo{author}{\bibfnamefont{D.~T.} \bibnamefont{Gillespie}},
  \bibinfo{journal}{J. of Comput. Phys.} \textbf{\bibinfo{volume}{22}},
  \bibinfo{pages}{403} (\bibinfo{year}{1976}).

\bibitem[{\citenamefont{Gillespie}(1977)}]{gillespie1977}
\bibinfo{author}{\bibfnamefont{D.~T.} \bibnamefont{Gillespie}},
  \bibinfo{journal}{J. Phys. Chem.} \textbf{\bibinfo{volume}{81}},
  \bibinfo{pages}{2340} (\bibinfo{year}{1977}).

\bibitem[{\citenamefont{Alves et~al.}(2020{\natexlab{a}})\citenamefont{Alves,
  Alves, Macedo-Filho, Ferreira, and Lima}}]{tfaalves2020}
\bibinfo{author}{\bibfnamefont{T.~F.~A.} \bibnamefont{Alves}},
  \bibinfo{author}{\bibfnamefont{G.~A.} \bibnamefont{Alves}},
  \bibinfo{author}{\bibfnamefont{A.}~\bibnamefont{Macedo-Filho}},
  \bibinfo{author}{\bibfnamefont{R.~S.} \bibnamefont{Ferreira}},
  \bibnamefont{and} \bibinfo{author}{\bibfnamefont{F.~W.~S.}
  \bibnamefont{Lima}}, \bibinfo{journal}{Phys. Rev. E}
  \textbf{\bibinfo{volume}{00}}, \bibinfo{pages}{000}
  (\bibinfo{year}{2020}{\natexlab{a}}).

\bibitem[{\citenamefont{Pastor-Satorras
  et~al.}(2015)\citenamefont{Pastor-Satorras, Castellano, Mieghem, and
  Vespignani}}]{pastorsatorras2015}
\bibinfo{author}{\bibfnamefont{R.}~\bibnamefont{Pastor-Satorras}},
  \bibinfo{author}{\bibfnamefont{C.}~\bibnamefont{Castellano}},
  \bibinfo{author}{\bibfnamefont{P.~V.} \bibnamefont{Mieghem}},
  \bibnamefont{and}
  \bibinfo{author}{\bibfnamefont{A.}~\bibnamefont{Vespignani}},
  \bibinfo{journal}{Rev. Mod. Phys.} \textbf{\bibinfo{volume}{87}},
  \bibinfo{pages}{925} (\bibinfo{year}{2015}).

\bibitem[{\citenamefont{Boyd}(19730)}]{boyd1973}
\bibinfo{author}{\bibfnamefont{D.~W.} \bibnamefont{Boyd}},
  \bibinfo{journal}{Canad. J. Math.} \textbf{\bibinfo{volume}{25}},
  \bibinfo{pages}{303} (\bibinfo{year}{19730}).

\bibitem[{\citenamefont{Alves et~al.}(2020{\natexlab{b}})\citenamefont{Alves,
  Alves, Lima, and Macedo-Filho}}]{gaalves2020}
\bibinfo{author}{\bibfnamefont{G.~A.} \bibnamefont{Alves}},
  \bibinfo{author}{\bibfnamefont{T.~F.~A.} \bibnamefont{Alves}},
  \bibinfo{author}{\bibfnamefont{F.~W.~S.} \bibnamefont{Lima}},
  \bibnamefont{and}
  \bibinfo{author}{\bibfnamefont{A.}~\bibnamefont{Macedo-Filho}},
  \bibinfo{journal}{Phys. A} \textbf{\bibinfo{volume}{00}},
  \bibinfo{pages}{000} (\bibinfo{year}{2020}{\natexlab{b}}).

\bibitem[{\citenamefont{Watts and Strogatz}(1998)}]{wattsstrogatz1998}
\bibinfo{author}{\bibfnamefont{D.~J.} \bibnamefont{Watts}} \bibnamefont{and}
  \bibinfo{author}{\bibfnamefont{H.}~\bibnamefont{Strogatz}},
  \bibinfo{journal}{Nature} \textbf{\bibinfo{volume}{393}},
  \bibinfo{pages}{440} (\bibinfo{year}{1998}).

\bibitem[{\citenamefont{Doye and Massen}(2005)}]{jspkdoyde2005}
\bibinfo{author}{\bibfnamefont{J.~P.~K.} \bibnamefont{Doye}} \bibnamefont{and}
  \bibinfo{author}{\bibfnamefont{C.}~\bibnamefont{Massen}},
  \bibinfo{journal}{Phys. Rev. E} \textbf{\bibinfo{volume}{71}},
  \bibinfo{pages}{016128} (\bibinfo{year}{2005}).

\bibitem[{\citenamefont{Shimoni et~al.}(2011)\citenamefont{Shimoni, Nudelman,
  Hayot, and Sealfon}}]{Shimoni-2011}
\bibinfo{author}{\bibfnamefont{Y.}~\bibnamefont{Shimoni}},
  \bibinfo{author}{\bibfnamefont{G.}~\bibnamefont{Nudelman}},
  \bibinfo{author}{\bibfnamefont{F.}~\bibnamefont{Hayot}}, \bibnamefont{and}
  \bibinfo{author}{\bibfnamefont{S.~C.} \bibnamefont{Sealfon}},
  \bibinfo{journal}{PLOS One} \textbf{\bibinfo{volume}{6}},
  \bibinfo{pages}{e29298} (\bibinfo{year}{2011}).

\bibitem[{\citenamefont{L\"{u}beck and Willmann}(2002)}]{lubeck2002}
\bibinfo{author}{\bibfnamefont{S.}~\bibnamefont{L\"{u}beck}} \bibnamefont{and}
  \bibinfo{author}{\bibfnamefont{R.~D.} \bibnamefont{Willmann}},
  \bibinfo{journal}{J. Phys. A: Math. Gen.} \textbf{\bibinfo{volume}{35}},
  \bibinfo{pages}{10205} (\bibinfo{year}{2002}).

\bibitem[{\citenamefont{Janssen et~al.}(2007)\citenamefont{Janssen, L\"{u}beck,
  and Stenull}}]{jansen2007}
\bibinfo{author}{\bibfnamefont{H.-K.} \bibnamefont{Janssen}},
  \bibinfo{author}{\bibfnamefont{S.}~\bibnamefont{L\"{u}beck}},
  \bibnamefont{and} \bibinfo{author}{\bibfnamefont{O.}~\bibnamefont{Stenull}},
  \bibinfo{journal}{Phys. Rev. E} \textbf{\bibinfo{volume}{76}},
  \bibinfo{pages}{041126} (\bibinfo{year}{2007}).

\bibitem[{\citenamefont{Henkel et~al.}(2008)\citenamefont{Henkel, Hinrichsen,
  and L\"{u}beck}}]{henkel2008}
\bibinfo{author}{\bibfnamefont{M.}~\bibnamefont{Henkel}},
  \bibinfo{author}{\bibfnamefont{H.}~\bibnamefont{Hinrichsen}},
  \bibnamefont{and}
  \bibinfo{author}{\bibfnamefont{S.}~\bibnamefont{L\"{u}beck}},
  \emph{\bibinfo{title}{Non-Equilibrium Phase Transitions}},
  vol.~\bibinfo{volume}{I} (\bibinfo{publisher}{Springer Science + Business
  Media B.V.}, \bibinfo{address}{Dordrecht}, \bibinfo{year}{2008}).

\end{thebibliography}

\end{document}